\documentclass[showpacs,preprint,aps]{revtex4}%
\usepackage[dvips]{graphicx}
\usepackage{epsfig}
\usepackage{psfrag}
\usepackage{amsmath}
\usepackage{amsfonts}
\usepackage{amssymb}
\usepackage{bm}
\usepackage{longtable}

\begin{document}


\title{One Dimensional Confinement of Electric Field and Humidity Dependent DNA Conductivity} 
\author{
{John M. Leveritt III, Carmen Dibaya, Sarah Tesar, Rajesh Shrestha,   \& Alexander L. Burin}
\\
\em
{$\dag$ Department of Chemistry, Tulane University, New Orleans, LA 70118,}\\
}
\newpage

\begin{abstract}
The dependence of DNA assemblies conductance on relative humidity is investigated theoretically. Following earlier suggestions we consider the ionic conductivity through the layers of water adsorbed by DNA molecules. The increase of humidity results in a growing water layer.  The binding energy of ions depends on the thickness of the water layer due to change in water polarization. This dependence is very strong at smaller thicknesses of water layers due to the low-dimensional confinement of an electric field in water. We show that the associated change in ion concentration can  explain the $6$ orders of magnitude increase of conductivity, with relative humidity growing from $0.05$ to $0.95$.  
\end{abstract}
\maketitle%

\section{Introduction}
\label{sec:intr}

Since the first experimental demonstration of charge transfer\cite{Barton} DNA molecules are considered promising candidates for the realization of nanodevices \cite{David,Porath,DNACond1,DNACond2,prl00,BBR}. Extensive investigation of molecular junctions made of long DNA molecules has resulted in many interesting, but sometimes controversial observations.  One common feature discovered in different experimental setups with long DNA molecules \cite{DNA1,DNA2,DNA3,DNA4,DNA5,DNA6,DNA7} is the dramatic sensitivity of dry DNA conductivity to the humidity of air. Indeed, the increase of humidity from $5\%$ to $95\%$ raises the conductivity of DNA molecular assemblies by six orders of magnitude.

The nature of this strong conductance dependence of relative humidity has been discussed by various authors \cite{DNA1,DNA2,DNA3,DNA4,DNA5,DNA6,DNA7}. It was suggested \cite{DNA2} to interpret observations in terms of ionic conduction through the water layers around DNA molecules. According to Refs. \cite{DNA2,DNA5,DNA7}, ions H$_{3}$O$^{+}$ and $OH^{-}$,  formed in water self-ionization processes, are responsible for observed conductivity. The binding energy of water depends on the relative humidity because the adsorption of water increases water solubility. Alternatively, there are theoretical models suggesting major electronic mechanism of conductivity (see e. g. Refs. \cite{Cuniberti,ch}). In our opinion, ions are the instruments of conduction rather than electrons.  According to Ref. \cite{ten} the activation energy of dry DNA for the carrier hopping through DNA duplex is smaller than that of k-
DNA in buffer, which contradicts with the increase of conductance with increasing relative humidity.
Also according to experiments \cite{DNA5,DNA6} single and double stranded DNA do not show a dramatic difference in conductance. In case of electronic mechanism of conductivity this result conflicts with the absence of charge transfer in a single stranded DNA \cite{Schuster}.  It is important to note that DNA conductance behaves differently in short DNA junctions \cite{DNA8}, where the conductivity is most likely associated with the electron transport.

\begin{figure}[ptbh]
\centering
\includegraphics[width=8cm,height=3cm]{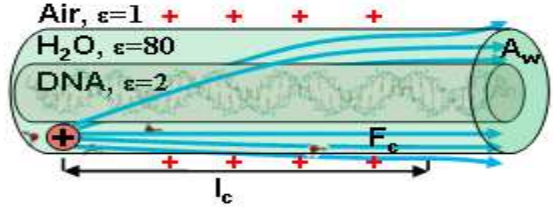}
\caption{Field confinement in the water tube formed around DNA in a humid environment (see text for details).}
\label{fig:DNA}
\end{figure}

In this manuscript we develop an electrostatic model to investigate the effect of humidity on ionic conductivity of DNA assemblies.
We treat each DNA molecule independently and assume each is surrounded by a layer of water Fig. \ref{fig:DNA}. In Sec. \ref{sec:II} the humidity dependent thickness of the water layer is estimated using experimental data  \cite{exp,exp1,Lavalle,exp3} and each dielectric layer in our model is characterized by its bulk dielectric constant.  Since dielectric constant of water ($\epsilon_{W}=80$) is much larger than that of the air ($\epsilon_{o}=1$) the electrostatic interaction in the thin water layer is different from that in the bulk water. The boundary conditions for the electric field between two dielectrics (water and air) results in the field confinement inside the water layer strongly affecting ion binding energies. Field confinement means that the electric field of interacting ions in water is substantially trapped in the water layer, and this effect causes DNA conductivity to be dramatically sensitive to humidity changes. The field confinement has been studied earlier for ionic channels \cite{Ref1,Ref2,Shklovskii1,Shklovskii2}, where we borrowed this idea. The field confinement results in stronger attraction between positive and negative ions.  Accordingly, fewer water molecules between ions results in an increase in binding energy and a decrease in conductance.  In Section \ref{sec:III} we calculate the  binding energy associated with electric field confinement using the exact solution of Poisson equations in the cylindrical geometry (Sec. \ref{sec:App}).

In Sec. \ref{sec:IV} the humidity dependence of ionic conductance associated with electric field confinement within the water layer is calculated and compared with existing experimental data. We obtained  an excellent agreement between theory and experiment in spite of the approximate character of our electrostatic model. 
A method for experimental verification of our model is to measure the non-linear voltage dependent conductance as proposed in Sec. \ref{sec:V}. We predict that a relatively small electric field $F_{c} \sim 10^{6}$V/m, such that $F_{c}l_{c}\sim k_{B}T$ is needed to sufficiently reduce the ion binding energy to produce a non-linear current-voltage dependence.

\section{Model}
\label{sec:II}

Consider a partially hydrated DNA molecule shown in Fig. \ref{fig:DNA}. One can distinguish dielectric cylindrical layers of DNA, water and air characterized by their dielectric constants $\epsilon_{in}$ (DNA), $\epsilon_{W}$ (water) and $\epsilon_{o}$ (air), respectively, and by radii  $r_{in}$ (DNA) and $r_{o}$. The radius of DNA and  bulk dielectric constants of air and water are known ($r_{in}\approx 1.3\AA$ in an A-conformation, $\epsilon_{o}\approx 1$ and $\epsilon_{W}\approx 80$). The dielectric constant of DNA is not well known because of the obvious difficulty of its direct measurement.  Different values for $\epsilon_{in}$ are proposed in different works. A low value for DNA dielectric constant $\epsilon \sim 2 - 4$ was suggested in Ref. \cite{eps2_4} which is typical of other organic polymers. Further experimental and theoretical studies suggest the dielectric constant of DNA is larger:  $\epsilon \sim 7$ \cite{eps71,eps72,eps73}, $\epsilon \sim 12.4 - 20$ \cite{eps12}, $\epsilon \sim 100$ \cite{eps100}.  We used different values for $\epsilon_{in}$ and found that the intermediate value $4\leq \epsilon_{in} \leq 12.4$ leads to  an agreement of our theory with experimental data for humidity dependent DNA conductance.

The humidity dependent radius of the water layer  $r_{o}$ is found using the experimental results for the number of adsorbed water molecules per nucleotide, $N^{ads}_{W}$ (see Table 1 in \cite{exp}, Fig. 1 in Ref. \cite{exp1}, \cite{Lavalle}).  It is important to note the adsorption measurements ignore the zeroth layer of water which contains approximately $6$ molecules per base pair even at zero humidity \cite{exp1,Lavalle,exp3}. These water molecules are bound to DNA due to electrostatic forces and/or hydrogen bonds.  The total number of water molecules per base pair can then be defined as $N_{W}\approx 2N^{ads}_{W}+6$.  Alternatively, the number of adsorbed water molecules can be described by the Branauer-Emmett-Teller (BET) equation, $N_{W}^{ads}=44rh(1-rh)^{-1}(1+19rh)^{-1}$ \cite{BET,Hill}, where $rh$ is relative humidity.  In our study, we use the experimental relationship because results from the BET equation can deviate from real behavior either if the water layer thickness approaches atomic size ($rh < 0.1$) or is comparable to the distance separating different DNA molecules $rh \rightarrow 1$.  The experimental data has no such limitations.  The humidity dependence of $N_{W}$ is summarized in Table \ref{tab}.

Using the number of water molecules per nucleotide $N_W$ at different relative humidity we can calculate the area of the water tube as $A_{W}=N_W(v_W/d_{in})$  and its radius as $r_{o}=\sqrt{A_{W}/\pi-r_{in}^{2}}$, where  $v_W=29.6\AA^3$ is the volume per single water molecule and $d_{in}=2.4\AA$ is the period of an A-DNA chain. The period of the DNA chain is the average distance between base pairs, which varies with DNA conformation. We  consider A-DNA because according to Ref. \cite{Lavalle}, DNA remains in A-conformation at humidity below $93\%$.  The radius of the water layer is larger and the binding energy is lower in A-DNA compared to B-DNA, so in the case of coexisting phases the formation of conducting ions will be more efficient in A-DNA.  Correspondingly, we use A-DNA radius $r_{in} \approx 1.3$nm for the inner cylinder in Fig. \ref{fig:DNA}.

It is not obvious whether the continuous medium approach is applicable to the ultrathin water layers containing few molecular layers (see Table \ref{tab}). The electrostatic model is more justified if we have at least two or more molecular layers of water, which takes place at relative humidities exceeding $0.5$. Since our model provides a good description of experimental data at all humidities we believe that it can be extended to lower humidities as well. One should notice that the recent modeling of dielectric relaxation of cytochrome $c$ oxidase \cite{Alexei} shows that even the  ultrasmall hydrophobic cavity around the catalytic center
in cytochrome $c$ can yet be characterized by the bulk water dielectric constant $80$. This gives an additional justification for our approach.

\begin{table}
\begin{tabular}
[c]{|l|l|l|l|l|l|l|l|l|l|l|l|l|l|l|l|}\hline
$rh$ & $0.05$ & $0.11$ & $0.23$ & $0.33$ & $0.39$ & $0.49$ & $0.54$ & $0.59$ & $0.65$ & $0.75$ & $0.80$ & $0.84$ & $0.87$ & $0.92$ &$0.93$\\\hline
$N_{W}$, exp. &  $8.64$ &    $9.36$  &    $10.6$  &    $12.1$  &    $12.7$  &  $14.5$  &  $    15.4$  &  $    16.7$  &  $   18.2 $  &  $   22.9 $  &  $   26.3 $  &  $   27.1   $  &  $ 32.2 $  &  $   46.1$ 
  &  $    47.5 $ \\\hline $A_{W}$, exp. &
  $1.08$ &    $    1.17 $ &    $   1.33$ &    $    1.51$ &    $    1.59$ &    $    1.8$ &    $    1.92$ &    $    2.07$ &    $    2.26 $ &    $   2.84 $ &    $   3.28 $ &    $   3.37 $ &    $   4.02$ &    $    5.74 $ &    $   5.91$ \\\hline $r_{o}$, exp. &
 $1.43$ &    $1.44$  &  $1.45$  &  $1.47$  &  $1.48$  &  $1.5$  &  $1.52$  &  $1.53$  &  $1.55$  &  $1.61$  &  $1.65$  &  $1.66$  &  $1.72$  &  $1.88$  &  $1.89$  \\\hline
$N_{W}$, theor. \cite{Hill} &  $8.38$  &  $    9.53$  &  $    10.9$  &  $    12.0$  &  $    12.7$  &  $    14.2$  &  $    15.2$  &  $   16.4$  &  $    18.3$  &  $
 23.3$  &  $   28.1$  &  $    33.2$  &  $38.6$  &  $    60.8$  &  $   70.5$  \\\hline
$A_{W}$, theor. &  $1.04$  &  $1.19$  &  $1.36$  &  $1.49$  &  $1.58$  &  $1.76$  &  $1.89$  &  $2.04$  &  $2.28$  &  $2.9$  &  $3.5$  &  $4.14$  &  $4.81$  &  $7.57$  &  $8.78$  \\\hline
$r_{o}$, theor. & $1.42$  &  $1.44$  &  $1.46$  &  $1.47$  &  $1.48$  &  $1.5$  &  $1.51$  &  $1.53$  &  $1.55$  &  $1.62$  &  $1.67$  &  $1.73$  &  $1.79$  &  $2.02$ &   $2.12$   \\\hline
\end{tabular}
\caption{Humidity ($rh$) and water layer parameters $N_{W}$, $A_{W}$ (nm$^{2}$) and $r_{o}$ (nm).}
\label{tab}
\end{table}


\section{Ion binding energy and field confinement}
\label{sec:III}

Following the previous discussion \cite{DNA2,DNA3,DNA4,DNA5,DNA6,DNA7} we investigate the ionic conductivity associated with $H_{3}O^{+}$ and  $OH^{-}$ ions formed in the water self-dissociation, which is referred to below as "pure water" conductance. The conductance depends on the density of ions, their mobilities, area of the water layer, and the length/ number of DNA molecules. We expect that the ion binding energy possesses the most significant humidity dependence because ion density depends on this energy exponentially. Only exponential dependence can be responsible for producing the six orders of magnitude change in humidity dependent DNA conductance. The conductance can be approximately expressed through the ion binding energy $U_{B}$ as \cite{DNA7}
\begin{equation}
C\propto \exp(-U_B/(2k_{B}T)).	
\label{eq:cond}
\end{equation}

The binding potential energy of two ions can be separated into two parts 
\begin{equation}
U_B=U_W+U_c.	
\label{eq:condbindenergy}
\end{equation}
The first contribution is the direct Coulomb attraction  which leads to the binding energy in bulk water, $U_{W}$. For instance the binding energy of OH$^{-}$ and H$^{+}$ ions forming H$_{2}$O molecule is given by $U_{W}~0.48$eV \cite{inorgtext}.  The second contribution to $U_{B}$ is is due to ion interaction with the induced polarization of dielectric environment at interfaces.  In the electrostatic model the humidity effect is fully related to the interaction with induced surface charges (see Sec. \ref{sec:App}) which create their own contribution to the binding energy $U_C$  absent in the bulk medium ($r_0\rightarrow \infty$). Changes of humidity modify  the thickness of water layer, $r_{o}$, and thus affects interaction of ions with dielectric polarization. 
The conductance dependence of humidity can be expressed as 
\begin{equation}
C=C_{0} \exp(-U_c/(2k_{B}T)),	
\label{eq:condf}
\end{equation}
 where $C_{0}$ is the preexponential factor describing the conductance in the limit when the radius of water layer approaches infinity ($U_{c}\rightarrow 0$). The binding energy $U_{c}$  depends only on the ion charges only because it is determined by the induced Coulomb interaction at distances larger than its ionic size (see Sec. \ref{sec:App}) . Therefore humidity dependence should be similar for different mechanisms of ion formation including water self-dissociation and cation escape from the charged phosphate group (see discussion in Sec. \ref{sec:IV}).

If, in accordance with \cite{DNA2,DNA7}, the conductance is due to  OH$^{-}$ and H$^{+}$ ions formed by self-dissociation of H$_{2}$O, then the  factor $C_{0}$ is determined by the conductivity of a pure bulk water ($c_{W}\sim 5.5\cdot 10^{-6}$Cm$/$m). The preexponential factor in the "pure water" conductance of the junction Eq. (\ref{eq:condf}) made of $N$ DNA molecules of the length $l$ connecting two electrodes can be estimated  as $C_{0}=c_{W}A_{W}N/l$ so the conductance of the whole system can be expressed as 
\begin{equation}
C=c_{W}\frac{A_{W}N}{l}\exp(-U_c/(2k_{B}T)),	
\label{eq:condf_total}
\end{equation}
Unfortunately, we did not find any information in experimental literature about the number of DNA molecules forming the junction and therefore we are unable to compare theoretical prediction for the absolute value of the "pure water" conductance with the experiment. 


The humidity dependent binding energy $U_c$ can be expressed as the work to bring a negative charge, initially located in the direct vicinity of a positive charge, to an infinite distance along the cylinder axis.  Due to the large difference between the dielectric constant of water and those of neighboring layers, the electric field is substantially confined inside the water layer \cite{Shklovskii1}.  Indeed, since the normal component of electric field, $E_n$, on the border of two dielectrics satisfies the condition $\epsilon_{W}E_{nW}=\epsilon_m E_{nm}$ ($m=Air$ or $DNA$) \cite{electrtext}, we can set $E_{nW}\approx 0$ in the limit $\epsilon_{m}/\epsilon_{W}\rightarrow 0$.  In this case, the force lines of electric field (blue arrows in Fig. \ref{fig:DNA}) do not escape from the water tube until the distance between the charge and the reference point is greater than the field confinement length $l_{c} < r_{o}$, thus representing the size of the surface charge domain.  The electric field, $F(r)$, within the domain restricted by the confinement length ($r<l_c$) can be estimated using the Gauss theorem;  the outward flux of an electric field through a closed surface ($F_c A_W$) is equal to $4\pi kq/\epsilon$ where $q=e$ is the charge of the region inside the surface.  This yields the following equation
\begin{equation}
F(r)=F_c\sim 2\pi ke/(\epsilon_{W}A_{W}).
\label{eq:conf_field}
\end{equation} 
This distance independent electric field is caused entirely by the surface charges which prevent the escape of field lines from the water layer.  Therefore, it describes the contribution to the binding energy related to DNA hydration that we are looking for.  It is important to note this field vanishes in the bulk limit $A_W \rightarrow$ corresponding to $100\%$ relative humidity, where the binding energy is determined solely by direct Coulomb forces.

The confinement length can be estimated using the continuity of the electric field tangential component at the boundary between two dielectrics and the field representations at confinement length as either the internal field, $F_c\sim 2\pi ke/(\epsilon_{W}A_{W})$, extrapolated from shorter lengths ($r<l_c$)  or the Coulomb field of a point charge, $F_C$=$ke/l_c^2$, extrapolated from longer lengths ($r>l_c$).  This yields $l_c\sim \sqrt{\epsilon_W A_{W}/(2\pi)}$ .  Finally, the energy to separate charges can be estimated as  $U_{c}=eF_cl_c\approx ke^2\sqrt{2\pi}/\sqrt{\epsilon_W A_{W}}$.  $A_{W} \sim 1$nm$^{2}$ (see Table \ref{tab}), we can estimate the confinement length as $l_{c} \sim 4$nm (accurate calculations result in a value one order of magnitude larger, see  Fig. \ref{fig:FieldConf}).  This size is much larger than all characteristic molecular sizes, which gives additional justification for the validity of our electrostatic approach.  Using our estimate for the ion binding energy associated with field confinement, we can predict the linear conductance logarithm dependence on the inverse square root of the water layer face area (cf. Eq. (\ref{eq:condf_total})) 
\begin{eqnarray}
\log(C)=\log(C_{0})-\eta \frac{\sqrt{2\pi}ke^{2}}{\sqrt{2\epsilon_W A_{W}}K_{B}T\log(10)}
\nonumber\\
=log(C_{0})-\eta S_{A}/A_{W}^{-1/2},  
\label{eq:C}
\end{eqnarray} 
where $\eta \sim 1$ is the proportionality constant between the estimate and the exact solution and $S_{A}=7.76$nm$^{-1}$.

\begin{figure}[ptbh]
\centering
\includegraphics[width=8cm]{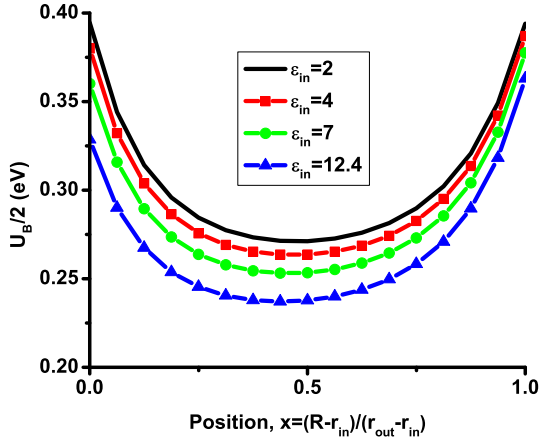}
\caption{Confinement energy at different positions of dissociating (water) molecule at relative humidity $0.75$.}
\label{fig:SzDep}
\end{figure}

\begin{figure}[ptbh]
\centering
\includegraphics[width=8cm]{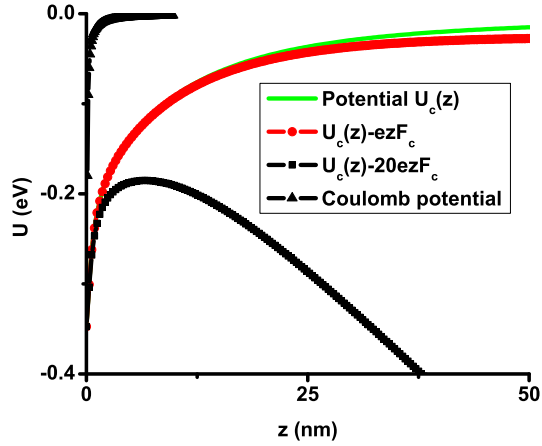}
\caption{Point charge potential dependence on the distance; $rh$=0.5. Induced field clearly exceeds the Coulomb field at distances exceeding few \AA. Electric field $F_{c}=2.5\cdot10^{5}$ V/m reduces the potential barrier for water self-dissociation by the thermal energy $k_{B}T \approx 0.026$eV while the field $10^{7}$V/m destroys confinement effect.}
\label{fig:FieldConf}
\end{figure}

Our estimate is very close to exact solutions of Poisson equations in cylindrical geometry given in Sec. \ref{sec:App}, where the direct Coulomb interaction and confinement binding energy are separated quite naturally.  These solutions are obtained using textbook methods \cite{electrtext}.  The solution for binding energy can be expanded over modified Bessel functions of the first and second kind, $I_{m}$ and $K_{m}$, as 
\begin{eqnarray}
U_B(R) = \frac{2ke^2}{\pi \epsilon_{W}}\sum_{m=0}^{\infty}(2-\delta_{m0})
\int_{0}^{\infty}dk(\Psi_{m}^{I}I_{m}(kR)+\Psi_{m}^{K}K_{m}(kR)); ~ \Psi_{m}^{I,K}=\frac{A^{I,K}}{D};
\nonumber\\
A^{I}= (\epsilon_{W}-\epsilon_{in})(\epsilon_{o}-\epsilon_{W})
K_{m}(kR)I_{m}(kr_{in})I_{m}'(kr_{in})K_{m}(kr_{o})K_{m}'(kr_{o}) 
\nonumber\\
- (\epsilon_{W}-\epsilon_{o})I_{m}(kR)K_{m}(kr_{o})K_{m}'(kr_{o})
(\epsilon_{in}I_m(kr_{in}) K_m'(kr_{in})
-\epsilon_{W}I_m'(kr_{in})K_m(kr_{in}));
\nonumber\\
A^{K}= (\epsilon_{W}-\epsilon_{in})(\epsilon_{o}-\epsilon_{W})
I_{m}(kR)I_{m}(kr_{in})I_{m}'(kr_{in})K_{m}(kr_{o})K_{m}'(kr_{o}) 
\nonumber\\
- (\epsilon_{W}-\epsilon_{in})K_{m}(kR)I_{m}(kr_{in})I_{m}'(kr_{in})
(\epsilon_{o}I_m(kr_{o}) K_m'(kr_{o})-\epsilon_{W}I_m'(kr_{o})K_m(kr_{o}));
\nonumber\\
D= (\epsilon_{W}-\epsilon_{in})(\epsilon_{o}-\epsilon_{W})
I_{m}(kr_{in})I_{m}'(kr_{in})
K_{m}(kr_{o})K_{m}'(kr_{o})
\nonumber\\
-(\epsilon_{in}I_m(kr_{in})K_m'(kr_{in})-\epsilon_{W}I_m'(kr_{in})K_m(kr_{in}))
(\epsilon_{o}I_m(kr_{o})K_m'(kr_{o})-\epsilon_{W}I_m'(kr_{o})K_m(kr_{o}))
\label{eq:BesselSolution}
\end{eqnarray}
Binding energies of ions Eq. (\ref{eq:BesselSolution}) are calculated using the Matlab software package.  This energy depends on the position $R$ of the positive charge (dissociating water molecule) with respect to the cylinder axis, $r_{in}<R<r_{o}$.  The binding energy has a minimum at the intermediate value of radius $R=R_{*}$  (Fig. \ref{fig:SzDep}) and tends to infinity at the layer boundaries, a characteristic of  the point charge model.  Since water splitting is most efficient for molecules at the binding energy minimum separated from both boundaries by distances exceeding the water molecule size, the point charge approach is justified.  Thus we used Eq. (\ref{eq:cond})  with binding energy $U_{B}(R_{*})$  to compare our theoretical predictions with the experimental data.  In the domain of interest $R \approx R_{*}$, the series of Bessel functions (Eq. (\ref{eq:BesselSolution})) rapidly converges.  Use of $10-20$ terms in expansion is sufficient to provide $1\%$ accuracy in calculations.

The humidity dependence of the binding energy is caused by the related dependence of the outer radius $r_{o}$ given by Table \ref{tab}. This radius is determined by humidity; it tends to zero if relative humidity approaches $0$ and goes to infinity when the relative humidity is close to $1$. Using this dependence we calculate the humidity dependence of conductance.

Note the confinement energy does not depend on the chemical nature of ions formed in dissociation, but rather on their charges.  Therefore, the formation of conducting ions by dissociation of any molecule can be characterized by the same  humidity dependence of conductance.  

To describe the field confinement effect, we also consider the distance dependent potential energy of two charges.  Both charges are located at distance $R_{*}$ from the polar axis and azimuth angle $\phi=0$, and separated by distance $z$.  The potential energy interaction related to confinement can be expressed similarly to Eq. (\ref{eq:BesselSolution})  as
\begin{eqnarray}
U_B(R) = \frac{2ke^2}{\pi \epsilon_{W}}\sum_{m=0}^{\infty}(2-\delta_{m0})
\int_{0}^{\infty}dk(\Psi_{m}^{I}I_{m}(kR_{*})+\Psi_{m}^{K}K_{m}(kR_{*}))cos(kz). 
\label{eq:BesselSolution1}
\end{eqnarray}
This expression was calculated using Matlab software for relative humidity $0.5$ and the results are shown in Fig. \ref{fig:FieldConf}.  It is clear from the graph the confinement related part of the interaction dominates at long distances of $25-50$nm;  therefore, the confinement length exceeds $10$nm. Since the associated potential barrier length for charge separation is very long, this barrier is extremely sensitive to the external electric field so strong non linearity in current-voltage dependence can be expected.  This effect is discussed later in \ref{sec:V}.

\begin{figure}[ptbh]
\centering
\includegraphics[width=8cm]{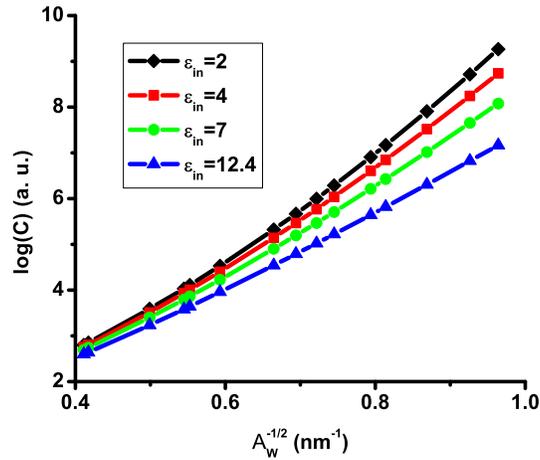}
\caption{Dependence of conductance on the area of water layer.}
\label{fig:ArDep}
\end{figure}

The dependence of conductance on the water layer area (Eq. (\ref{eq:cond})) for different DNA dielectric constants is shown in Fig. \ref{fig:ArDep}.  The calculated behavior agrees with the qualitative prediction (Eq. (\ref{eq:C})).  Moreover, the slopes of the linear conductance dependence on $A_{W}^{-1/2}$ also agree with Eq. (5) Eq. (\ref{eq:C}) (Table \ref{tab2}), which proves the significance of an electric field confinement.

\section{Comparison of theory with experimental data}
\label{sec:IV}

\begin{figure}[ptbh]
\centering
\includegraphics[width=8cm]{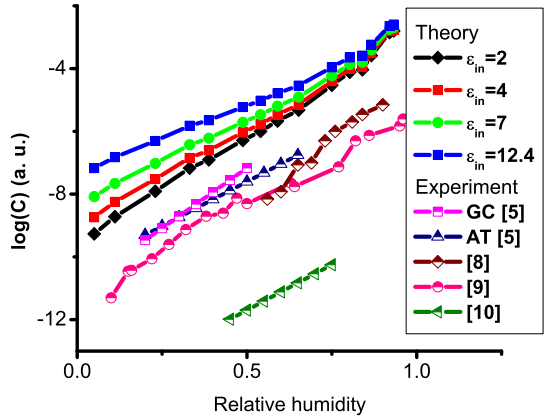}
\caption{Experimental data and theoretical predictions for humidity dependent conductance.}
\label{fig:HDep}
\end{figure}

Combining calculations of conductance dependence on water layer area (Fig. \ref{fig:ArDep}) and experimental data for area dependence on relative humidity (Table \ref{tab}), one can describe the humidity dependence of conductance.  This dependence is shown in Fig. \ref{fig:HDep} for different dielectric constants of DNA ($\epsilon_{in}$) together with the available experimental data. We were not able to predict the pre-exponential factor; therefore, all conductance values are given in arbitrary units.  

It follows from Fig. \ref{fig:HDep} that the calculated conductance shows exponential dependence on humidity as in experiments; moreover its behavior agrees quantitatively with majority of the experimental data for $4< \epsilon_{in}<12$.    The theoretical and experimental slopes of exponential conductance dependence on humidity are similar as shown in Table \ref{tab2}.  We can see if the DNA dielectric constant is set to $7$ as suggested in Refs. \cite{eps71,eps72,eps73}, then we obtain an exponential dependence slope $d\ln C/d rh \approx 0.58$.  This value is identical to the one observed in several studies including poly-A – poly-T DNA \cite{DNA2}, single and double stranded herring DNA \cite{DNA6}, and $\lambda$-DNA \cite{DNA7}.  In other experiments with poly-G – poly-C DNA \cite{DNA2} and DNA linked gold particles \cite{DNA5}, larger slopes ($0.7$ and $0.9$) have been reported.  Much stronger dependence has been observed for DNA modified with acridine orange \cite{DNA3} which is most likely due to its strong chemical difference from standard DNA.  Our electrostatic model can fit all of this data (Table \ref{tab2}).

Note that the slopes can be sensitive to  fluctuations in the bundle structure supporting water dissociation.  The obvious example of such fluctuation is the overlap of two adjacent DNA molecules or the water layers formed around them.  In this structure, the water layer face area $A_W$ will be approximately doubled (cf. Fig. \ref{fig:DNA}).  This doubling reduces the humidity dependent part of binding energy by a factor of $\sqrt{2}$ ((\ref{eq:C})), which also reduces the slope of the conductance exponential dependence by the same factor.  The overall effect of fluctuations depends on their probabilities which is not known to us, but can vary in different experiments.  The fluctuations will lead to the weakening of humidity dependence and the experimental data characterized by the weakest dependence could have the strongest fluctuation effect.

In addition to the humidity dependence, other interesting properties of DNA conductance at various humidity were investigated, including the sequence 
and temperature dependencies \cite{DNA2,DNA5},  and comparison of humidity dependent conductance of single and double stranded DNA \cite{DNA6}.  
The experimental temperature dependence of DNA conductance has been well approximated by the Arrhenius law.  In Ref. \cite{DNA2}, the temperature dependence is reported at relative humidity $0.5$ and the conductivity activation energy has been estimated as $0.49$eV for a poly-G – poly-C sample and $0.6$eV for poly-A – poly-T sample.  We can estimate the ion transport activation energy, $E_{A}$, for "pure water conductance" using Eqs. (\ref{eq:cond}), (\ref{eq:condbindenergy}) and our calculations.  For DNA dielectric constants $\epsilon_{i}= 2, 4, 7, 12.4, 20$ we get, $E_A=0.62,0.6,0.58,0.55,0.52$eV,  respectively.  All estimates agree reasonably well with experimental values.

In Ref. \cite{DNA5}, the activation energy was studied for large relative humidities, $0.7$ and $0.85$.  It was found to be less than that of pure water conductance which is $0.24$eV.  Therefore, the formation of ions in this system is most likely associated with a process other than the self-dissociation of water molecules.  Yet this process seems to also be sensitive to the field confinement and humidity in good agreement with our theory.  Perhaps the process is the breaking of a bond between the ion and phosphate group which has smaller bulk binding energy than that of the water molecule.

The humidity dependent conductances in single and double stranded herring DNA \cite{DNA6}  are found similar to each other.  This is not what is expected in our model.  According to Ref. \cite{exp1}, the number of adsorbed water molecules per nucleotide is approximately the same for both single and double stranded DNA. Accordingly, the total number of water molecules per unit length and area of the water layer, $A_W$, is smaller for single stranded DNA by a factor of $2$. Since the confinement energy Eq. (\ref{eq:C}) is inversely proportional to $A_{W}^{-1/2}$  the binding energy associated with the field confinement should increase by the same factor of $\sqrt{2}$ and so does the slope of the humidity dependence. This expectation conflicts with the experimental observation of nearly identical slopes for single and double stranded DNA.  How can this discrepancy be explained?  It is possible the conductance was measured in an essentially non-linear regime in Ref. \cite{DNA6} (current voltage dependence has been reported at bias voltage $4$V where nonlinearity is obviously significant).  Our theory is only applicable to the linear regime.  One should notice the four-contact measurements in Refs. \cite{DNA2,DNA5} are performed in the linear regime, while nonlinearity is clearly seen in the current-voltage characteristics observed in Refs. \cite{DNA2,DNA5}.  Perhaps the small value of slope and the identical behavior of single and double stranded DNA conductance is the consequence of non-linear regime of measurements.  Also, the two contact method was used in both Refs. \cite{DNA6,DNA7}.  The conductance measurements made using this method are sensitive to the junction resistance, which is excluded in four contact measurements \cite{DNA2,DNA5}.










\begin{table}
\begin{tabular}
[c]{|l|l|l|l|l|l|}\hline
$\epsilon_{in}$ & $2$ & $4$ & $7$ & $12.4$ & $20$
\\\hline
$\eta=\frac{dlog(C)}{S_{A}d A_{W}^{-1/2}}$, theory  &  $1.49 $ &    $   1.39$ &    $     1.25$ &    $     1.06$ &    $     0.87$ 
\\\hline
Slope $\frac{dlog(C)}{d rh}$, theory  &  $0.7 $ &    $    0.64$ &    $     0.58$ &    $     0.49$ &    $     0.4$ \\\hline
Experiment & \cite{DNA2} & \cite{DNA2}& \cite{DNA5} & \cite{DNA6} & \cite{DNA7}\\\hline
Details & GC &  AT & 
DNA linked gold & herring DNA & $\lambda$-DNA\\\hline
Slope ($rh$), experiment  &  $0.76$ &    $    0.58$ &    $     0.9$ &    $     0.58$ &    $     0.58$ \\\hline
\end{tabular}
\caption{Conductance dependence on the area of water layer and humidity}
\label{tab2}
\end{table}

\section{Discussion and conclusions}
\label{sec:V} 

Thus we suggested the solution of the long-standing problem of the humidity dependence of DNA conductance. The dramatic increase in DNA conductance of six orders of magnitude with increasing humidity is interpreted using the change of ion binding energy induced by the electric field confinement within the water layer. This confinement is caused by the large value of the water dielectric constant compared to the environment. Consequently the binding energy of conducting ions is very sensitive to the thickness of water layer determined by humidity. Increasing the humidity results in a growing water layer, which reduces the field confinement. Consequently, attraction force between two oppositely charged ions decreases and  activation energy of ionic transport decreases accordingly leading  to a dramatic increase of ionic density and conductivity. This behavior is opposite to the one expected for electronic transport where activation energy is expected to grow with the amount of water surrounding DNA \cite{ten}. 
 
Yet depending on particular experiments, there are problems in interpretation of conductance mechanisms due to strong variations in conductance temperature dependence  \cite{DNA6} and the absence of denaturation effect of conductance \cite{DNA5}.  Therefore, it would be interesting to perform direct experimental verification of field confinement.  We believe that the measurements of a non linear current-voltage characteristic can be used as such verification.  Strong nonlinearity of current-voltage characteristics of DNA is expected because the confinement length is much longer than characteristic molecular sizes (Fig. \ref{fig:FieldConf}). 
Indeed the external DC electric field $F$ should reduce ion binding energy and support ion dissociation in the relatively large domain of field confinement restricted by confinement length $l_{c}$ (see Fig. \ref{fig:FieldConf}).  Assuming this DC field is much smaller than the confinement field $F_{c}$ Eq. (\ref{eq:conf_field}) one can describe its effect as a perturbation.  Then we can estimate the reduction of ion binding energy as $\delta U_{c}= -Fl_{c}$.  The nonlinear conductance can be approximately represented as
\begin{equation}
C(F) =C(0)\exp\left(\frac{Fl_{c}}{2k_{B}T}\right). 
\label{eq:nonlin_cond}
\end{equation}
The nonlinearity should become strong  at $F_{c}\sim 2k_{B}T/l_{c}$.  According to our solution for ion binding energy distance dependence (Eq. (\ref{eq:BesselSolution1}), we can estimate $F_{C}\sim 8\times 10^{5} V/m$ for relative humidity $0.5$.  The threshold field is similar for other values of humidity.  The measurements of this dependence (Eq. (\ref{eq:nonlin_cond})) are very interesting because they result in the direct determination of the confinement length.  One should notice the four contact measurements in Ref. \cite{DNA2}  were made at DC fields less than $8.3\cdot 10^{5}$V/m.  The nonlinear threshold for those measurements can be higher than our prediction because DNA molecules were aligned under some angle to the field.  In Ref. \cite{DNA5}, the maximum electric field was near $3\cdot 10^{5}$V/m. We suggest an increase of the field by one order of magnitude which should lead to remarkable nonlinearity according to our model.  The application of a field exceeding $20 F_c$ should remove the confinement effect as illustrated in Fig. \ref{fig:FieldConf}. The exponential field dependence should become weaker when the DC field approaches the confinement field $F_{c}$. Thus a non-linear current voltage characteristic can be used to extract both confinement field and confinement length that can be compared to our electrostatic model.

{\it Note added in proof}. While our paper was under review,
another paper \cite{38} considering humidity dependent conductivity
of DNA appeared in press. This paper suggests an explanation
of experimental data using two-dimensional field confinement
in planar water layers surrounding DNA films.
Their model is both quantitatively and qualitatively different
than our model which uses one-dimensional field confinement.
We believe that further experimental investigation of
humidity dependent conductivity, particularly in the nonlinear
regime, will help choose which explanation is more
relevant.

\section{Acknowledgment}
This work is supported by the NSF CRC Program, Grant No. 0628092. The authors acknowledge Boris Shklovskii for informing us about the problem, Fred Lewis for useful discussion and informing us about the controversy around DNA dielectric constant and Torsten Fiebig, Russ Schmehl, Igor Rubtsov and Gail Blaustein for useful discussions.

\section{Appendix} 
\label{sec:App}

Here we derive the exact solution for the potential of a point charge located in the water layer between DNA and air Fig. \ref{fig:DNA} and express the ion binding energy using that potential.  The charge $e$ is placed in water and its cylindrical coordinates are chosen as  ${\bm r}_{0}=(\rho, 0, 0)$, $r_{in}<\rho<r_{o}$. The charge potential satisfies the Laplace equation
\begin{equation}
\Delta \Phi = 4\pi k \frac{e}{\epsilon}\delta({\bm r}-{\bm r}_{0}),
\label{eq:Poisson}
\end{equation}
and the standard boundary conditions of continuity of potential in the border separating different media and discontinuity of its normal derivatives
\begin{equation}
\epsilon_{1}\frac{\partial \Phi_{1}}{\partial n_{1}}= \epsilon_{2}\frac{\partial \Phi_{2}}{\partial n_{2}}. 
\label{eq:PoissonBoundary}
\end{equation}
Using the cylindrical symmetry of the problem the solution can be expanded into the series of modified Bessel functions following the textbook  \cite{electrtext}. Inside the DNA (${\bm r}=(\rho, z, \varphi)$) one should use modified Bessel functions of the first kind having no singularities at $\rho=0$.   The solution at the point $0<r<r_{in}$ can be expressed as
\begin{equation}
\Phi_{in}({\bm r})=\frac{2ke}{\pi \epsilon_{W}}\sum_{m=0}^{\infty}(2-\delta_{m0})\int_{0}^{\infty}dk \Phi_{m}^{in}(k)I_{m}(k\rho)cos(kz)cos(m\varphi).  
\label{eq:BesselIn}
\end{equation}
where $\delta_{m0}$ is the Kronecker symbol and functions $ \Phi_{m}^{in}(k)$ are unknowns to be determined. The solution in the air $r_{o}<r$ can be expressed similarly with the only difference that one should use modified Bessel functions of the second kind approaching zero when their argument tends to infinity 
\begin{equation}
\Phi_{out}({\bm r})=\frac{2ke}{\pi \epsilon_{W}}\sum_{m=0}^{\infty}(2-\delta_{m0})\int_{0}^{\infty}dk \Phi_{m}^{out}(k)K_{m}(k\rho)cos(kz)cos(m\varphi).  
\label{eq:BesselOut}
\end{equation}
The solution inside the water layer can be expressed as the superposition of the field of point charge satisfying Eq. (\ref{eq:Poisson}) and the general solution of Laplace equation using modified Bessel functions of both first and second kind (cf. Ref. \cite{electrtext}) 
\begin{eqnarray}
\Phi_{W}({\bm r})=\frac{2ke}{\pi \epsilon_{W}}\sum_{m=0}^{\infty}(2-\delta_{m0})
\nonumber\\
\times
\int_{0}^{\infty}dk \left(K_{m}(k\rho_{0})I_{m}(k\rho)\theta(\rho_{0}-\rho)+K_{m}(k\rho)I_{m}(k\rho_{0})\theta(\rho-\rho_{0})\right)cos(kz)cos(m\varphi)
\nonumber\\
+\frac{2ke}{\pi \epsilon_{W}}\sum_{m=0}^{\infty}(2-\delta_{m0})\int_{0}^{\infty}dk \left(\Psi_{m}^{I}(k)I_{m}(k\rho)+\Psi_{m}^{K}(k)K_{m}(k\rho)\right)cos(kz)cos(m\varphi).  
\label{eq:BesselW}
\end{eqnarray}
The first term represents the Coulomb field of the point charge in the dielectric medium, which is independent of the system geometry and describes the case of an infinite homogeneous sample.  The second term is the surface charge contribution which describes the geometry effect and correspondingly, the humidity dependence of interest.  This contribution to the binding energy is determined by the potential difference between the bound state,  $\Phi_{W}({\bm r}_{0})$ (the position of the second charge coincides with the position of the first), and the state where the second charge is infinitely far from the first,  $\Phi_{W}(\infty)=0$. Thus the confinement induced binding energy for ions having charges $e$ and $-e$ can be expressed as 
\begin{eqnarray}
U_{c}=\frac{2ke^2}{\pi \epsilon_{W}}\sum_{m=0}^{\infty}(2-\delta_{m0})\int_{0}^{\infty}dk \left(\Psi_{m}^{I}(k)I_{m}(k\rho)+\Psi_{m}^{K}(k)K_{m}(k\rho)\right).  
\label{eq:AnsBindEn}
\end{eqnarray}
Unknown expansion coefficients $\Psi_{m}^{I,K}$ can be determined using boundary conditions of potential continuity and electric field discontinuity Eq. (\ref{eq:PoissonBoundary}) at the boundaries.  These expressions along with the binding energy Eq. (\ref{eq:BesselSolution}) are given within the main body of the present text.


\end{document}